The new Geological Age that never was or the multiple layers of the Transientocene

# The new Geological Age that never was or the multiple layers of the Transientocene

*Orfeu Bertolami*
Departamento de Física e Astronomia,
Faculdade de Ciências, Universidade do Porto;
Centro de Física das Universidades do Minho e do Porto
orfeu.bertolami@fc.up.pt

**Abstract**
Since its humble origins, humans have left imprints on the face of the planet. From the profound transformation unleashed by the Neolithic Revolution, about 12000 years ago, till the present, humans have reshaped the planet significantly. From the second half of the XX century, the impact on the atmosphere, biosphere, cryosphere, hydrosphere and upper lithosphere is so overwhelming that a new geological age, the Anthropocene, was proposed to consider the extent of these transformations. However, despite the ubiquitous nature of the changes in course, the International Union of Geological Sciences rejected in March 2024 formalizing the Anthropocene as a new geological epoch. This controversial decision implies that geologists are not quite convinced that human activities have reached the level of an encompassing new geological age. Nevertheless, it is beyond any doubt that there is no single spot on the planet where the signs of the transformations ensued by the human activities are not felt. Furthermore, the interconnection of the human activities has reached a level of entanglement that it makes the Anthropocene an inescapable feature of our present and immediate future. Thus, more important than framing our present condition in a way that it can be recognised by geologists in the future, is the understanding that by its very nature, the Anthropocene is a condition that is continuously being reshaped to the point that we should instead regard our time as a Transientocene, a time of significant and multidimensional transformations.

**Keywords:** Anthropocene; Climate crisis; Crises and Polycrisis; Transientocene.

**Resumo**
Desde os primórdios, as atividades humanas deixaram marcas específicas na face do planeta. Das profundas transformações desencadeadas pela Revolução Neolítica que teve lugar há cerca de 12000 anos até o presente, as atividades humanas atingiram proporções globais e incontornáveis. Desde a segunda metade do século passado, o impacto na atmosfera, biosfera, criosfera, hidrosfera e litosfera superior é tão significativo que uma nova era geológica, o Antropocénico, foi proposta para caracterizar a extensão das transformações em curso. Contudo, a União Internacional das Ciências Geológicas rejeitou, em Março de 2024, a formalização desta proposta. Esta controversa decisão implica que os geólogos não estão convencidos de que a ação humana no seu conjunto equivale a uma nova era geológica, embora seja inegável que os efeitos desta se façam sentir em qualquer parte do planeta. Há que considerar que a conectividade das ações humanas atingiu um ponto de entrelaçamento que o Antropocénico e, a problemática a que está associado, é um aspeto incontornável do nosso presente e que será determinante para o nosso futuro. Assim, para além de compreender os múltiplos desdobramentos do Antropocénico, parece-nos mais relevante pensar o nosso tempo como um contínuo multidimensional de indeléveis transformações, o Transientocénico.

"The ants Geiser recently observed under a dripping fir tree are not concerned with what anyone might know about them; nor were the dinosaurs, which died out before a human being set eyes on them. All the papers, whether on the wall or on the carpet, can go. Who cares about the Holocene? Nature needs no names. Geiser knows that. The rocks do not need his memory."

Max Frisch, *Man in the Holocene*

The new Geological Age that never was or the multiple layers of the Transientocene

## 1. The Anthropocene is dead. Long live the Anthropocene!

The evolution of the human activities till its equivalence to a geological force is a development that has been, since its inception, fundamentally transformative. Humans shape the landscape and continuously seek expanding the capacity of getting the most from the territory they settle in. In fact, given its fragile framework and the severity of the environment, the humble hordes of hunters and gatherers that managed to survive the harshness of Glaciation periods that peaked at 25000 and 19000 years ago, thanks to the climatic stability of the Holocene, manage to multiply their population and accumulate the necessary skills to initiate farming activities, initially at the Fertile Crescent, and China at about 12000 years ago and somewhat later in Americas. The ubiquity and independence of these activities show the success and the universality of the human expansion impulse. An extraordinary mixture of necessity and curiosity has led humans to modify the environment till becoming the dominant species on the planet, reaching space and building weapons of mass destruction in the second half of the last century and, more recently, devising intelligent algorithms that can surpass their own capacity to solve complex problems. It was just inevitable that this expansion would put a mounting pressure on the natural resources of the planet and would have an indelible impact on the components of the Earth System (atmosphere, biosphere, cryosphere, hydrosphere and upper lithosphere), whose interconnected performance has allowed the planet to host an extraordinary exuberance of life forms.

The drastic changes on the climate patterns, the massive loss of biodiversity and the widespread of environmental pollution abundantly show that the Earth System is no longer being driven only by natural drivers (astronomical, geological and internal), but instead, is under the dominant force of the human activities (Crutzen & Stoermer 2000; Crutzen 2002). This shift, which is felt globally from the second half of the last century onwards (Steffen at al 2015) and on the oceans somewhat from 1970’s onwards (Jouffray et al 2020), has prompted the concept of a new geological epoch, the Anthropocene, to follow the geologically well characterized Holocene epoch, whose climatic stability was particularly important for the flourishing of the human civilisations.

It is just natural that such a proposal would be thoroughly scrutinized by the International Commission on Stratigraphy (ICS) and the International Union of Geological Sciences (IUGS). In 2016, the Anthropocene Working Group (AWG) was created to proceed towards a golden spike proposal based on potential chronostratigraphic markers, which led to 12 candidate sites to be selected. The purpose was to stablish a representative stratigraphic profile of new epoch and to define its onset. The search process culminated in July 2023 with the proposal that the sediments of the Crawford Lake in the Halton Region in Canada were a suitable representative of the Great Acceleration, the post-World War II period in which global population, economic growth, massive exploitation of natural resources and the emergence of the atomic age took place.

However, in 2024, after more than a decade of deliberation, the AWG proposal was outvoted by a wide margin by the Subcommission on Quaternary Stratigraphy of which the AWG was part of. Subsequently, ICS and the IUGS confirmed, almost unanimously, the rejection of the AWG's Anthropocene Epoch proposal as a Geologic Time Scale, even though it acknowledged that the Anthropocene would remain to be used by Earth scientists as well as by social scientists, politicians and economists as an invaluable marker of the human impact on the Earth System. Hence, despite its rejection on strictly chronostratigraphic scientific grounds, the Anthropocene was recognised to be a quite useful conceptual landmark, invaluable for understanding the extent of the changes the Earth System is going through since 1950’s. Furthermore, it is widely perceived that the transitional nature of the current evolution of the Earth System and its implications for our collective future cannot be fully appreciated without the conceptual framework provided by the Anthropocene.

## 2. The Age of Polycrisis, Novacene or Transientocene

Indeed, the impact of the human activities is so profound that it is consensual among the experts that it is pushing the Earth System towards a new equilibrium state away from the climatic equilibrium of the Holocene. This new equilibrium state is usually referred to as the Hothouse Earth state (Steffen et al., 2018) and is shown to be an attractor of trajectories in the phase space of a thermodynamical model of the Earth System (Bertolami & Francisco 2019). If so, the Anthropocene is necessarily a transitional stage of the Earth System from the Holocene to the Hothouse Earth state. In fact, the possible range of trajectories of the Earth System during the Anthropocene potentially admits quite hazardous evolutions, which include bifurcations and chaotic behaviour depending on the rate of growth of the human activities, implying that the behaviour of the Earth System can, under conditions, become harder to predict (Bernardini, Bertolami & Francisco, 2022).

In fact, the objective state and evolution of the Earth System can be captured by the so-called Planetary Boundaries (Rockström et al., 2009; Steffen et al., 2015), a set of at least nine parameters (climate change, biosphere integrity, land-system change, biogeochemical flows, ocean acidification, freshwater change, pollution, atmospheric aerosol loading, stratospheric ozone depletion), which frame the state of the Earth System with respect to its standing during the Holocene. Furthermore, it can be established empirically that the Planetary Boundaries are, on their own, highly connected and do interact in a non-negligible way with each other (see e.g. Barbosa, Bertolami & Francisco, 2020). In fact, it is proposed that these interactions can be used to potentially boost resilience of the Earth System (Bertolami & Nyström 2026). At present, all nine Planetary Boundaries have been screened and seven of them (Richardson et al., 2023; Sakschewski et al., 2025) have already gone through what is regarded as the Holocene outer boundary, the so-called Safe Operating Space. This means that in concrete terms, humankind is heading towards the unknown and most likely the future under the current conditions will imply profound changes in the main ecosystems of the planet (Steffen et al., 2018) and unfolding of climate disasters with dire implications for millions of people. Therefore, it is no longer possible to dismiss the climate crisis, which is manifesting itself ubiquitously, leaving behind a visible and traumatic trail of destruction and suffering. Indeed, this climatic crisis and the problematic times we are living have been dubbed as the age of the polycrisis, a concept first discussed in the context of the social sciences (Morin & Kern 1999), but that was recently revived in order to emphasize that a single crisis can lead to a profusion of entangled crises and impact on the Earth System in harmful ways (Lawrence et al., 2024; Søgaard Jørgensen et al., 2024). The importance of understanding the entanglement of crises and how they evolve and affect the Earth System has prompted us to devise a model inspired on matrix models, considered in condensed matter physics and string theory, to couple social-political events to the Planetary Boundaries so to yield a predictive dynamical model (Bertolami & Elísio 2026). Once fully developed, these models can potentially allow for objective comparisons on the effect on the Planetary Boundaries of different socio-political events, which might be a relevant guide for considering decisions to face risks and measures of adaptation and mitigation[1].

Furthermore, it is important to realise that the rate of change imposed by human activities on the Earth System is so much faster than the natural time scales that it puts in question whether a static depiction of these transformations via a single chronostratigraphic profile is useful in geological terms. This might be at the core of the problem that geologists faced when considering the decision to bind themselves to a chronostratigraphic representation of the Anthropocene, which by its very nature, might be too ephemeral to be reliably framed in this way. In this respect, it might be sensible to consider that the Anthropocene cannot be fully represented by a single chronostratigraphic profile, but instead by a series of profiles that depict the most encompassing and representative changes on the technological developments that drive human activities. Indeed, it is predictable that the debris related to the hardware of computers that have grown exponentially after the information-internet revolution at the last decades of the last century will leave an indelible profile on the upper lithosphere. Likewise, the debris of renewable energy devices will also accumulate and leave a distinct and identifiable imprint. It is just natural to assume an increase on the presence of rare metals on representative stratigraphic profiles in the next decades and so on. Of course, these future stratigraphic profiles will be indicative of the specific transformations of their time even if the increasing concern of protecting the environment of their hazardous effects ask for setting up specific conditions to store them.

---

[1] In what concerns adaptation and mitigation measures, we defend that a more structured approach should be considered and a *universal resilience social tax* upon consumption should be implemented (see e.g. Bertolami 2022). The economical context of this measure was discussed (Bertolami & Gonçalves 2023), leading to new parameter for the assessment of risks, namely, *safety* (Bertolami & Gonçalves 2025). Of course, these measures aim to address the broader issue of internalisation of the natural capital (see e.g. Bertolami & Francisco 2021; Bertolami 2024b).

For sure, a crucial landmark in the recent development of the human civilisation is the emergence of the Internet and the digitalization era. This development was recently discussed in the context of its effects on the Earth System (Bertolami 2025) and will not be repeated here. Furthermore, another undeniable new landslide event is the emergence of Artificial Intelligence (AI) (see also Bertolami 2025, for a discussion on the impact on the Earth System). James Lovelock, the author and scientist that called our attention to the idea that the complex feedback mechanisms and behaviour of the Earth System are fundamentally regulated by its biota and has to be regarded as a whole, very much like a living being, Gaia (Lovelock 1972; Lovelock & Margulis 1974; Lovelock 1995; 2000; 2009), considered this development a foundational historical event. Indeed, he has stressed, in a final contribution as an active voice of wisdom, that the emergence of AI is so unique in human history that its singularity deserves being properly acknowledged. He suggested that this coming age of hyperintelligence should be referred to as Novacene (Lovelock 2019). Lovelock regarded the transformations engendered by the Novacene with moderate optimism, if suitable regulatory measures are to be implemented.

For sure, the Novacene poses a whole set of challenges, from the societal point of view to the specific implications to the climate crisis itself. Indeed, the implications of replacing the human labour force by robotic AI driven entities (AIDEs) has such a multitude of implications that it is hard even to conceive. In fact, even if one assumes that such a transition can occur without major social disruption, it is inevitable that it requires redirecting the energetic resources from the real world to the virtual one so to feed the widespread use of AI and the massive army of AIDEs that will replace human labour. These potentially perilous developments were previously discussed (see e.g. Bertolami 2024a; Bertolami 2025; Mendes 2025) and their problematic nature must be considered in the context of the web of issues that get entangled within an Anthropocene in continuous acceleration. Common sense demands that the development of AI must be strongly regulated and its social impact in the process of replacing the human labour must be counterbalanced. A minimal set of measures should include devising a suitable set of taxation rules and parity rates when replacing human labour by AIDEs as well as setting up a sensible quota on the amount of energy AIDEs can withdraw from the global energy budget. Of course, it is urgent that these matters are properly addressed and widely discussed.

In any case, given the inherent fleetingness of the Anthropocene, we suggest that the current stage of the human society is referred to as Transientocene (or Impermanentocene), the age in which the acceleration of events and their evanescence are the only features one can take for granted in an endless chain of continuous transformations. Thus, as the density of meaningful changes encompassed by the Transientocene is so high, they cannot be captured by a single representative chronostratigraphic profile. In fact, the Transientocene challenges the core of what one understands by a geological time scale. Nevertheless, in our understanding, the transformations engendered by the current human actions leave a permanent and non-trivial imprint in the Earth System and hence on the upper lithosphere that is of direct concern of geologists at present and in the future. Therefore, if a single chronostratigraphic profile is inadequate to characterize the Anthropocene, theoretically, the Transientocene can only be fully depicted by a whole set of multilayer chronostratigraphic profiles. Only their whole set can represent the Transientocene.

It is worth stressing that the Transientocene, challenges our very capacity of describing and capturing the complexity of the current chain of developments. This overflow of changes suggests that the clock of historical evolution should be reset to a new and ultrafast time scale (see also Chakrabarty 2009, Mendes 2025). We should no longer aim to obtain a correspondence between the human actions and the geological forces, but instead acknowledge that human civilisation, and the new developments such as the new form of intelligence that is emerging from our mastering of the information technology is a qualitatively new chapter whose impact on society and on the Earth System must be urgently gauged and properly framed in order to avoid a further acceleration of the social problems and of the climate crisis. It is worth mentioning that the everchanging nature of the market coupled to quick technological and scientific developments intensifies the frailty of the individuals and that has already been discussed in the social sciences. This inherent instability on the standing of the individuals in society has been identified as a distinctive feature of what was dubbed as the hypermodernity times (Lipovetsky 2004).

In fact, the Transientocene can also be seen as an abrupt discontinuity of scales. The ultra-fast and ephemeral time of the human actions and socio-political events are imposing transformations on the Earth System that were, till the Great Acceleration, within the realm of natural phenomena whose time spans were dictated by the laws of Natures. One can further stress this mismatch considering that beyond the eons of geology there is the cosmological time scale and the fact that the Universe engendered by the Big Bang, 13.8 billion years ago, must admit self-conscious observers. Indeed, any description and modeling of the Universe must admit biophysical and chemical conditions that are causally compatible with the existence of self-aware cosmological observers. This necessity, often referred to as the Weak Anthropic Principle (see Barrow & Tipler 1986, for an extensive discussion), implies that only a Universe that admits planets with Gaian features in which the biota can effectively act as an efficient regulator of environmental conditions, can host evolved self-aware, intelligent organisms that eventually become cosmological observers. A more radical form of this principle establishes that the existence of cosmological observers sets nontrivial constraints on the laws of physics and on the fundamental constants of Nature (see e.g. Fellgett 1988). Of course, these constraints concern our existence on a cosmological time scale, which is being endangered by the ultra-fast mounting anthropogenic runaway greenhouse effect and other transformations engendered in the Transientocene. These radical changes on the planetary conditions compromise the long-term sustainability of all species on the planet. Thus, if human actions jeopardise the future generations, shouldn't one consider them in terms of a collective responsibility and in an ethical context? Shouldn't one think in terms of a Cosmic Responsibility? (Bertolami 2010; Bertolami & Gomes 2018). Of course, these concepts are useless unless they can be internalised in a way that allow for reshaping the economic and political dimensions of the human activities. In any case, the implications are clear: the ultra-fast changes that characterize the Transientocene are a direct threat to the long-term sustainability of our species as well as on the existence of all others harboured by our planet.

## 3. The trumpets of alarm

The Transientocene calls for an urgent rethinking of the principles that have been guiding world's decision systems. Reacting to the urgent problems that continuously arise from the flow of events in a complex and highly connected world is no longer effective without a coherent strategy and a vision for the future. We are witnessing the derailing of the precarious, nevertheless tendentiously peaceful, consensus that was laboriously built after the World War II. Human civilisation is losing the capacity to manufacture the necessary stability to solve the problems that endanger its long-term existence.

Our present time is plagued with worrisome developments: the callous disregard for the regulating measures on the emission of greenhouse gases by many countries leading to gloomy prospects for the future; intensification of the confrontation between countries fuelled by the rebirth of fascism, populism and irrational nationalist agendas; the widespread use of misinformation and political and geopolitical violence; the breakdown of the rules of international law and of the mediation institutions that were, till recently, consensually accepted as conveyers of agreement, most particularly the United Nations. The list is much longer, but it does not need to be exhaustive neither to raise the alarms of fear and despair nor for the purpose the present discussion. Decisive action is required to address the mounting problems associated with the climate crisis, but above all, a thoughtful and long-term reflection is asked to restore the confidence in the cardinal values of the civilised world, through empowering the democratic channels of discussion and its methodology to reach consensus on human rights and on the principles of international law.

However, despite the obvious need of reflection, the pressure to reshape the old-world order by violence and war seems unstoppable. When the forces of destruction are so often the rhetoric and the code of action of those in power, one fears for the future of civilisation. One wonders, how it is possible that thousands of years of human civilisation have been insufficient to build the collective strength to overcome the tension between Eros (life) and Thanatos (death), identified by Freud as a basic ancestral instinct that guided all individuals. This should be a point of reflection. Technological developments have empowered us in unthinkable ways, but they have also provided us with the means to engender huge inequalities, destitution and collective annihilation. Furthermore, these developments did not percolate deeply enough into the human institutions to the point of changing our ancestral differences and prejudices. Most often, these developments lead to a rupture with procedures that were setup to avoid that human's groups and countries relapse into violence as a natural way to settle differences and other ancestral forms of conflict. These ruptures are becoming more often, which introduce an erratic pattern on the trajectories of the Transientocene. These chaotic developments are amplified by irrational impulses of the voluntarist nature of political leaders that lost the shame of embracing fascism and populism as their political guidelines. These trends will accelerate the climate crisis and lead humankind to the verge of a conflict of world proportions.

In what concerns the foreseeable AI revolution, it is likely that it might transcend, in its implications, the known historical trends of replacing an obsolete and less productive technology by a more performing new one. The AI revolution delivers not the only possibility of relieving us from perilous and unwanted work, but of replacing us altogether. Without a radical transformation of the guideline principles of the current economy and its prioritization of maximal profit, most of the humankind run the risk of being made redundant and replaced by a more performing and more malleable working force of AIDEs. Nothing could better to install this apocalyptic agenda than a humankind in disarray, divided by wars and irreconcilable political interests. A palpable fear is that the polycrisis age of a hyper fluid Transientocene opens the gates for radical social transformations which bring such an abrupt and apocalyptic replacement of the human working force by an army of AIDEs that society has no means to regulate and to build up the proper safeguards. Unfortunately, this undisclosed agenda is at the heart of most of the speeches about the virtues of AI and its impact on economy. It is hard not to think that the emphasis on the gains of productivity and its potential for freeing human from the displeasing work, is yet another neoliberal piece of rhetoric that aims to keep the existing controllers of the economic and political power in charge and capable of controlling the agenda of the forthcoming technological developments.

We are a successful species thanks to our capacity to collaborate in large numbers by trusting in fictional constructions that are valued collectively (see e.g. Harari 2024). However, some fictional constructions can overvalue differences and disregard common interests and fundamental rights. The fictions that claim that flattening educational and wealth inequalities is an obstacle to economic development must be refuted on theoretical as well as on concrete grounds. Solidarity does not hinder economic prosperity, not even in the short term.
To counterbalance a chaotic and erratic Transientocene guided by a profit oriented hyperintelligence, and the irrational and opportunistic behaviour of political leaders, one should bring back the universal values of humanism and solidarity. This involves replacing the deeply rooted competitive spirit built within the worldwide educational system by a new model based on collaboration. Compassion, empathy and social concern should be taught on equal footing as to writing, reading and acquiring basic mathematical skills. We have the duty of equipping the future generations with skills that will allow them to reshape the current events and to think in solutions that are more generous and broader minded. There is no shortage of pedagogical examples in this direction and studies that show that an encompassing change on the basic values of the current education paradigm is desirable and perfectly possible (see e.g. Arendt et al., 2000; Vygotsky, 1986). Our collective future cannot be stained by the pessimism of the powerless common citizen systematically exploited by the greed of an oligarchy of privileged individuals and politicians. History is rich in developments that engendered fairness and social justice. These foundational glimpses cannot be forgotten. The future is always an open set of alternatives, and we should keep in mind that implementing the benevolent aspects of our common culture is the most effective way to keep the flow of events open and compatible with a long-term sustainability of the future generations.

**Acknowledgments**
The author would like to thank João Ribeiro Mendes, Ricardo Pedrosa Elísio and Rivaldo França Mello for important discussions and suggestions.